\documentclass[12pt,usenames,dvipsnames]{article}

\usepackage{latexsym}
\usepackage{amssymb,amsfonts,amsmath}
\usepackage{graphicx} 
\usepackage{indentfirst}
\usepackage{bbm}
\usepackage{amssymb}
\usepackage{verbatim}
\usepackage{amsmath, amsthm,amssymb}
\usepackage{mathrsfs}
\usepackage{hyperref}
\usepackage{amsfonts}
\usepackage{dsfont}
\usepackage{cite}
\usepackage{xcolor}
\usepackage{enumerate}
\usepackage{cleveref}

\parskip=\medskipamount

\newcommand {\cE}{{\cal E}}

\newcommand {\cL}{{\cal L}}

\newcommand {\cN}{{\cal N}}

\newcommand {\cS}{{\cal S}}

\def\a{\alpha}

\def\b{\beta}

\def\d{\delta}

\def\l{\lambda}

\def\o{\omega}

\def\q{\theta}

\def\s{\sigma}

\def\z{\zeta}
\def\D{\Delta}
\def\F{\Phi}

\def\O{\Omega}

\def\U{\Upsilon}
\def\X{\Xi}

\def\rd{{\rm d}}
\def\ri{{\rm i}}
\def\re{{\rm e}}

\newcommand{\ad}{{\dot{\alpha}}}                           
\newcommand{\bd}{{\dot{\beta}}}                            
\newcommand{\ve}{\varepsilon}                            

\renewcommand{\aa}{{\a\ad}}
\newcommand{\bb}{{\b\bd}}
\newcommand{\pa}{\partial}                           
\newcommand{\hf}{\frac12}

\newcommand{\be}{\begin{equation}}
\newcommand{\ee}{\end{equation}}
\newcommand{\bea}{\begin{eqnarray}}
\newcommand{\eea}{\end{eqnarray}}
\newcommand{\non}{\nonumber}
\newcommand{\bm}[1]{\mbox{\boldmath$#1$}}

\def\double #1{#1{\hbox{\kern-2pt $#1$}}}

\newif\ifdtup

\newcommand{\bsubeq}{\begin{subequations}}
\newcommand{\esubeq}{\end{subequations}}

\numberwithin{equation}{section}

\newcommand{\sU}{\mathsf{U}}

\begin{document}

\begin{titlepage}
\begin{flushright}
August, 2023 \\
\end{flushright}
\vspace{5mm}

\begin{center}
{\Large \bf 
Self-duality for $\cal N$-extended superconformal gauge multiplets 
}
\end{center}

\begin{center}

{\bf Sergei M. Kuzenko and Emmanouil S. N. Raptakis} \\
\vspace{5mm}

\footnotesize{
{\it Department of Physics M013, The University of Western Australia\\
35 Stirling Highway, Perth W.A. 6009, Australia}}  
~\\
\vspace{2mm}
~\\
Email: \texttt{ 
sergei.kuzenko@uwa.edu.au, emmanouil.raptakis@uwa.edu.au}\\
\vspace{2mm}

\end{center}

\begin{abstract}
\baselineskip=14pt
We develop a general formalism of duality rotations for $\cal N$-extended superconformal gauge multiplets in conformally flat backgrounds as an extension of the approach given in arXiv:2107.02001. Additionally, we construct $\mathsf{U}(1)$ duality-invariant models for the ${\mathcal N}=2$ superconformal gravitino multiplet recently described in arXiv:2305.16029.
Each of them is automatically self-dual with respect to a superfield Legendre transformation. A method is proposed to generate such self-dual 
models, including a family of superconformal theories. 
\end{abstract}
\vspace{5mm}

\vfill

\vfill
\end{titlepage}

\newpage
\renewcommand{\thefootnote}{\arabic{footnote}}
\setcounter{footnote}{0}

\tableofcontents{}
\vspace{1cm}
\bigskip\hrule

\allowdisplaybreaks

\section{Introduction}

Since the pioneering work by Gaillard and Zumino \cite{GZ1}, the general formalism of duality rotations 
has been developed for nonlinear electrodynamics \cite{GR1,GR2,GZ2,GZ3}
and its $\cN=1$ and $\cN=2$ supersymmetric extensions
\cite{KT1,KT2}.\footnote{The main results of \cite{KT1,KT2} have been generalised to curved supergravity backgrounds \cite{KMcC,K12}.} (For a comprehensive review of these and related developments, see \cite{KT2,AFZ}.)
In order to generate such self-dual theories, auxiliary variable formulations have been developed for nonlinear electrodynamics\footnote{An auxiliary variable setting for self-dual nonlinear electrodynamics was also used in \cite{HKS}.}
   \cite{IZ_N3,IZ1,IZ2,IZ3}, as well as  for $\cN=1$ and 
$\cN=2$ supersymmetric duality-invariant theories coupled to supergravity  \cite{K13}
(in the $\cN=1$ rigid supersymmetric case,
analogous results were independently obtained in \cite{ILZ}).

Motivated by the discovery of the ModMax theory \cite{BLST} and  its supersymmetric counterpart \cite{BLST2,K21}, 
the formalism of $\sU(1)$ duality rotations has recently been extended in \cite{KR21-2}
to higher-spin conformal gauge fields on conformally flat backgrounds and some of their $\cN=1$ and $\cN=2$ superconformal cousins. Specifically, the following types of $\sU(1)$ duality-invariant dynamical systems were studied in \cite{KR21-2}:
\begin{itemize} 

\item Self-dual models for  a conformal field, $\phi_{\a(m) \ad(n)}$, $m,n \geq1$, with the gauge freedom \eqref{ComplexGT}, where $\nabla_{\a\ad} = (\s^a)_{\aa} \nabla_a$ is a conformally covariant derivative. Here $\nabla_a = e_{a}{}^m \partial_m - \hf \o_{a}{}^{bc} M_{bc}$, 
where $e_{a}^{m}$ and $\o_{a}{}^{bc}$ denote the vielbein and spin connection, respectively. The corresponding actions are functionals of the gauge invariant field strengths $\hat{\mathbb{C}}^{[\D]}_{\a(m+n)} $ and
$	\check{\mathbb{C}}^{[\D]}_{\a(m+n)}$, which are  defined by 
\eqref{ComplexFS} and have the conformal properties \eqref{dimensions}.

\item Self-dual models for an $\cN=1$ superconformal real prepotential $\U_{\a(s) \ad(s)} = \overline{\U_{\a(s) \ad(s)}}$, with $s>0$,  defined modulo the gauge transformations
\bea
\d_\z \U_{\a(s) \ad(s)} = \nabla_{(\a_1} \bar{\z}_{\a_2 \dots \a_s) \ad(s)} - \bar{\nabla}_{(\ad_1} \z_{\a(s) \ad_2 \dots \ad_s)}~,
\eea
where $ \nabla_A = (\nabla_a, \nabla_\a , \bar \nabla^\ad)$ are the $\cN=1$ superconformally covariant derivatives. The action of such a model is a functional of the gauge-invariant chiral field strength 
\bea
{\mathbb W}_{\a(2s+1)} = - \frac{1}{4} \bar{\nabla}^2 \nabla_{(\a_1}{}^{\bd_1} \dots \nabla_{\a_s}{}^{\bd_s} \nabla_{\a_{s+1}} \U_{\a_{s+2} \dots \a_{2s+1}) \bd(s)} ~, 
\qquad \bar \nabla^\bd {\mathbb W}_{\a(2s+1)} = 0
\eea
and its conjugate. The $s=0$ case corresponds to a vector multiplet, and the corresponding duality-invariant models for the $\cN=1$ vector multiplet were studied in  
\cite{KT1,KT2,KMcC}.

\item Self-dual models for an $\cN=2$ superconformal real prepotential $\U_{\a(s) \ad(s)}$, with $s\geq0$, defined modulo the gauge transformations
\begin{align}
	s\geq1 :& \qquad \d_\z \U_{\a(s)\ad(s)} = \nabla_{(\a_1}^i \bar{\z}_{\a_2 \dots \a_s) \ad(s)i} + \bar{\nabla}_{(\ad_1}^i \z_{\a(s) \ad_2 \dots \ad_s) i}  ~, 
	\\
	s=0 :& \qquad \d_\z \U = \nabla^{ij} \bar{\z}_{ij} + \bar{\nabla}_{ij} {\z}^{ij}  ~.
\end{align}
The action of such a model is a functional of the gauge-invariant chiral field strength
\bea
{\mathbb W}_{\a(2s+2)} =  \bar{\nabla}^4 \nabla_{(\a_1}{}^{\bd_1} \dots \nabla_{\a_s}{}^{\bd_s} \nabla_{\a_{s+1} \a_{s+2}} {\U}_{\a_{s+3} \dots \a_{2s+2}) \bd(s)} ~,
\qquad \bar \nabla^\bd_j {\mathbb W}_{\a(2s+2)} = 0~,
\label{1.4}
\eea
where we have introduced 
the chiral projection operator 
$\bar{\nabla}^4  := \bar{\nabla}^{ij} \bar{\nabla}_{ij}
$
and the second-order operators
$
\nabla_{\a\b} := \nabla_{(\a}^k \nabla_{\b) k} $, $ \bar{\nabla}^{\ad\bd} := \bar\nabla^{(\ad}_k \bar\nabla^{\bd) k}$, $\nabla_{ij} = \nabla^{\a (i} \nabla^{j)}_\a$ and $\bar{\nabla}^{ij} = \bar{\nabla}_\ad^{(i} \bar{\nabla}^{\ad j}$,
with  $\nabla_A = (\nabla_a, \nabla_\a^i , \bar \nabla^\ad_i)$ the  $\cN=2$ superconformal covariant derivatives. 
The $s=0$ case corresponds to the conformal supergravity multiplet; the corresponding  gauge transformation is given by eq. \eqref{SCGTc}.

\end{itemize} 
At the same time, $\cN$-extended superconformal gauge-invariant models have been constructed \cite{KR21} to describe the dynamics of a complex tensor superfield $\U_{\a(m)\ad(n)}$, with $m,n \geq 0$, in a conformally flat superspace. Such a prepotential is defined modulo the gauge transformations \eqref{SCGT}, and the corresponding gauge-invariant chiral field strengths  $\hat{\mathbb{W}}_{\a(m+n+\cN)}^{[\D]} $ and 
$\check{\mathbb{W}}^{[\D]}_{\a(m+n+\cN)} $ are given in \eqref{SUSYFS}. One of the main goals of this paper is to 
develop a general formalism of $\mathsf{U}(1)$ duality rotations for these $\cal N$-extended superconformal gauge multiplets in conformally flat backgrounds.\footnote{The formalism of duality rotations for $\cal N$-extended superconformal gauge multiplets described by a real prepotential $\U_{\a(s) \ad(s)}$, with $s \geq 0$, was recently described in \cite{ERThesis}.}

As follows from \eqref{1.4} and  \eqref{SUSYFS}, the chiral field strengths carry at least two spinor indices in the $\cN=2$ case. A chiral scalar field strength $\mathbb W$ is known to correspond to a vector multiplet \cite{GSW}. The only missing choice of chiral spinor field strengths, $\hat{\mathbb{W}}_{\a}$ and 
$\check{\mathbb{W}}_{\a} $,  has been shown to correspond to the $\cN=2$ superconformal gravitino multiplet discovered in \cite{HKR}. The second goal of this work is to construct duality-invariant models for this multiplet.

This paper is organised as follows. In section \ref{Section2}, we review the formalism of $\sU(1)$ duality rotations for conformal gauge fields. Then, in section \ref{Section3}, we extend this formalism to the case of $\cN$-extended superconformal gauge multiplets. Finally, section \ref{Section4} is devoted to the construction of $\sU(1)$ duality-invariant models for the $\cN=2$ superconformal gravitino multiplet. The main body of this paper is accompanied by a single technical appendix, appendix \ref{appendixB}. It is devoted to a reduction of the free action for the $\cN=2$ superconformal gravitino multiplet to $\cN=1$ superspace.

All results in this work are presented within the framework of conformal (super)space. In the non-supersymmetric case, we employ the approach of \cite{KTvN} recast 
in the modern setting of \cite{BKNT-M1}.
 For the $\cN=1$ and $\cN=2$ superconformal cases, we refer the reader to 
 the original publications \cite{ButterN=1} and \cite{ButterN=2}, respectively,
 as well as to  \cite{KRTM1, KRTM2} for a recent review, whose conventions we utilise. For $\cN>2$, our conventions coincide with the ones of \cite{KKR}. 

The two-component spinor notation and conventions we employ in this work follow \cite{BK}, which are similar to those of \cite{WB}. Additionally, throughout this paper we make use of the convention whereby indices denoted by the same symbol are to be symmetrised over, e.g. 
\begin{align}
	U_{\a(m)} V_{\a(n)} = U_{(\a_1 . . .\a_m} V_{\a_{m+1} . . . \a_{m+n})} =\frac{1}{(m+n)!}\big(U_{\a_1 . . .\a_m} V_{\a_{m+1} . . . \a_{m+n}}+\cdots\big)~, \label{convention}
\end{align}
with a similar convention for dotted spinor indices. 

\section{Conformal gauge fields}
\label{Section2}
This section is devoted to a review of the formalism of $\sU(1)$ duality rotations for conformal gauge fields developed in \cite{KR21-2}. We recall that such a field, $\phi_{\a(m) \ad(n)}$, with $m,n \geq1$, is defined modulo the gauge transformations \cite{KP,KMT,Vasiliev}
\be
\label{ComplexGT}
\d_\ell \phi_{\a(m)\ad(n)} = \nabla_{\aa} \ell_{\a(m-1) \ad(n-1)} ~,
\ee
where $\nabla_\aa = (\s^a)_{\aa} \nabla_a$ is a conformally covariant derivative. It should be pointed out that, for $m=n=s$, it may be consistently restricted to be real; $\overline{\phi_{\a(s) \ad(s)}} = \phi_{\a(s) \ad(s)}$.
Consistency of \eqref{ComplexGT} with conformal symmetry implies that $\phi_{\a(m)\ad(n)}$ is a primary field of dimension $2-\hf(m+n)$
\begin{align}
	K_\bb \phi_{\a(m) \ad(n)} = 0 ~, \qquad \mathbb{D} \phi_{\a(m) \ad(n)} = \Big(2 - \hf(m+n)\Big) \phi_{\a(m)\ad(n)} ~,
\end{align}
where $K^a$ and $\mathbb{D}$ denote the special conformal and dilatation generators, respectively.

From $\phi_{\a(m)\ad(n)}$, we construct the following field strengths
\begin{align}
	\label{ComplexFS}
	\hat{\mathbb{C}}^{[\D]}_{\a(m+n)} = (\nabla_\a{}^\bd)^n \phi_{\a(m) \bd(n)} ~, \qquad
	\check{\mathbb{C}}^{[\D]}_{\a(m+n)} = (\nabla_{\a}{}^{\bd})^m \bar{\phi}_{\a(n) \bd(m)} ~,
	\qquad \D = m-n~.	
\end{align}
In the 
$m=n\equiv s$ case, the field strengths \eqref{ComplexFS} coincide, $\hat{\mathbb{C}}^{[0]}_{\a(2s)} = \check{\mathbb{C}}^{[0]}_{\a(2s)}$. The descendants \eqref{ComplexFS} are primary in generic backgrounds, 
\begin{subequations} \label{dimensions}
	\bea 
	K_\bb \hat{\mathbb{C}}^{[\D]}_{\a(m+n)} &=&0~, \qquad 
	\mathbb{D} \hat{\mathbb{C}}^{[\D]}_{\a(m+n)} = \Big(2 - \frac \D 2\Big) \hat{\mathbb{C}}^{[\D]}_{\a(m+n)}~,\\
	K_\bb\check{{\mathbb{C}}}^{[\D]}_{\a(m+n)}&=&0~,\qquad \mathbb{D}\check{{\mathbb{C}}}^{[\D]}_{\a(m+n)}
	=\Big(2+\frac{\D}{2}\Big)\check{{\mathbb{C}}}^{[\D]}_{\a(m+n)}~.
	\eea
\end{subequations} 
and gauge-invariant in all conformally flat ones,
\bea
C_{abcd} =0 \quad \implies \quad \d_\ell \hat{\mathbb{C}}^{[\D]}_{\a(m+n)} = \d_\ell  \check{\mathbb{C}}^{[\D]}_{\a(m+n)} =0~.
\eea
Here $C_{abcd}$ denotes the background Weyl tensor.
In what follows, we will assume such a geometry.

It follows from \eqref{dimensions} that the dimensions of the field strengths \eqref{ComplexFS} are determined by $\D$, which explains why the field strengths carry the label $\D$.
We also note that the field strengths \eqref{ComplexFS} are related via the Bianchi identity
\be
\label{ComplexBI}
(\nabla^\b{}_\ad)^m \hat{\mathbb{C}}^{[\D]}_{\a(n) \b(m)} 
= (\nabla_\a{}^\bd)^n \bar{\check{\mathbb{C}}}_{\ad(m) \bd(n)}^{[\D]} ~.
\ee

\subsection{$\sU(1)$ duality rotations for conformal gauge fields}

We consider a dynamical system describing the propagation of $\phi_{\a(m) \ad(n)}$. The corresponding action functional, which we denote $\cS^{(m,n)}[\hat{\mathbb{C}},\check{\mathbb{C}}]$, is assumed to depend only on $\hat{\mathbb{C}}^{[\D]}_{\a(m+n)}$, $\check{\mathbb{C}}^{[\D]}_{\a(m+n)}$, and their conjugates. Considering $\cS^{(m,n)}[\hat{\mathbb{C}},\check{\mathbb{C}}]$ as a functional of the unconstrained fields $\hat{\mathbb{C}}^{[\D]}_{\a(m+n)}$, $\check{\mathbb{C}}^{[\D]}_{\a(m+n)}$ and their conjugates, we may introduce the primary fields
\be
\ri^{m+n+1} \hat{\mathbb{M}}^{[\D]}_{\a(m+n)} :=  \frac{\d \cS^{(m,n)}[\hat{\mathbb{C}},\check{\mathbb{C}}]}{\d \check{\mathbb{C}}^{{[\D]} \a(m+n)}} ~, 
\qquad \ri^{m+n+1} \check{\mathbb{M}}^{[\D]}_{\a(m+n)} :=  \frac{\d \cS^{(m,n)}[\hat{\mathbb{C}},\check{\mathbb{C}}]}{\d \hat{\mathbb{C}}^{{[\D]} \a(m+n)}}~,
\label{CFD}
\ee
where the variational derivative is defined in the following way
\begin{align}
	\d \cS^{(m,n)}[\hat{\mathbb{C}},\check{\mathbb{C}}] &= \int \rd^4x\, e \, 
	\Big \{ \d \hat{\mathbb{C}}^{{[\D]} \a(m+n)} \frac{\d \cS^{(m,n)}[\hat{\mathbb{C}},\check{\mathbb{C}}]}{\d \hat{\mathbb{C}}^{{[\D]} \a(m+n)}} 
	+ \d \check{\mathbb{C}}^{{[\D]},\a(m+n)} \frac{\d \cS^{(m,n)}[\hat{\mathbb{C}},\check{\mathbb{C}}]}{\d \check{\mathbb{C}}^{{[\D]} \a(m+n)}} \Big \} + \text{c.c.}~,
\end{align}
and $e$ denotes the determinant of the (inverse) vielbein. The conformal properties of the fields \eqref{CFD} are: 
\begin{subequations} 
	\bea 
	K_\bb \hat{\mathbb{M}}^{[\D]}_{\a(m+n)} &=&0~, \qquad 
	\mathbb{D} \hat{\mathbb{M}}^{[\D]}_{\a(m+n)} = \Big(2 - \frac \D 2\Big) \hat{\mathbb{M}}^{[\D]}_{\a(m+n)}~, \\
	K_{\bb} \check{{\mathbb{M}}}^{[\D]}_{\a(m+n)}&=&0~,\qquad \mathbb{D}\check{{\mathbb{M}}}^{[\D]}_{\a(m+n)}
	=\Big(2+\frac{\D}{2}\Big)\check{{\mathbb{M}}}^{[\D]}_{\a(m+n)}~.
	\eea
\end{subequations} 
Varying $\cS^{(m,n)}[\hat{\mathbb{C}},\check{\mathbb{C}}]$ with respect to $\bar{\phi}_{\a(n) \ad(m)}$ yields the following equation of motion
\bea
\label{ComplexEoM}
(\nabla^\b{}_\ad)^m \hat{\mathbb{M}}^{[\D]}_{\a(n) \b(m)} 
= (\nabla_\a{}^\bd)^n \bar{\check{\mathbb{M}}}^{[\D]}_{\ad(m) \bd(n)} ~.
\eea

It is clear from the discussion above that the system of equations \eqref{ComplexBI} and \eqref{ComplexEoM} is invariant under the $\sU(1)$ duality rotations
\begin{subequations}
	\label{ComplexU(1)}
	\begin{align}
		\d_\l \hat{\mathbb{C}}^{[\D]}_{\a(m+n)} &= \l \hat{\mathbb{M}}^{[\D]}_{\a(m+n)} ~, \quad \d_\l \hat{\mathbb{M}}^{[\D]}_{\a(m+n)} = - \l \hat{\mathbb{C}}^{[\D]}_{\a(m+n)}  ~, \\
		\d_\l\check{\mathbb{C}}^{[\D]}_{\a(m+n)} &= \l \check{\mathbb{M}}^{[\D]}_{\a(m+n)}~, \quad \d_\l \check{\mathbb{M}}^{[\D]}_{\a(m+n)} = - \l \check{\mathbb{C}}^{[\D]}_{\a(m+n)} ~,
	\end{align}
\end{subequations}
where $\bar{\l} = \l$ is an arbitrary constant parameter. One may then construct $\sU(1)$ duality-invariant nonlinear models for such fields.
They may be shown to satisfy the self-duality equation \cite{KR21-2}
\bea
\label{ComplexSDE}
\ri^{m+n+1} \int \rd^4x \, e \, \Big \{ \hat{\mathbb{C}}^{[\D] \a(m+n)}  \check{\mathbb{C}}^{[\D]}_{\a(m+n)}
+ \hat{\mathbb{M}}^{[\D] \a(m+n)} \check{\mathbb{M}}^{[\D]}_{\a(m+n)} \Big \} + \text{c.c.}  = 0 ~,
\eea
which must hold for unconstrained fields $\hat{\mathbb{C}}^{[\D]}_{\a(m+n)}$ and $\check{\mathbb{C}}^{[\D]}_{\a(m+n)}$.
The simplest solution of this equation is the free action \cite{FT,FL,FL2,KMT,KP}
\bea
\cS^{(m,n)}_{\rm Free}[\hat{\mathbb{C}},\check{\mathbb{C}}]
= {\ri^{m+n}}\int \rd^4 x \, e\,  \hat{ {\mathbb{C}}}^{[\D] \a(m+n)}\check{{\mathbb{C}}}^{[\D]}_{\a(m+n)} 
+{\rm c.c.}
\eea

The $m=n=1$ case in \eqref{ComplexSDE} corresponds to nonlinear electrodynamics where the integral form of the self-duality equation was given for the first time in \cite{KT2}. In the earlier publications \cite{GR1,GZ2,GZ3,B-B} it was given in the form
\bea
G^{ab}\, \tilde{G}_{ab} +F^{ab} \, \tilde{F}_{ab} =0~,\qquad 
\tilde{G}_{ab} (F):=
\hf \, \ve_{abcd}\, G^{cd}(F) =
2 \, \frac{\pa L(F)}{\pa F^{ab}}~,
\label{NLED}
\eea
where $L(F)$ is the Lagrangian of the electromagnetic field. 
As emphasised in \cite{KT2}, the integral form of the self-duality equation must be used in theories with higher derivatives. 
Duality-invariant theories 
with higher derivatives were studied, e.g., in \cite{AFZ,Chemissany:2011yv,AF,AFT}.


\subsection{Self-duality under Legendre transformations}
\label{section2.2}

In the case of nonlinear (super) electrodynamics,  $\sU(1)$ duality invariance implies self-duality under Legendre transformations, see \cite{KT2} for a review and \cite{KR21-2} for the extension to bosonic higher spins. In this subsection, we will show that this property extends to the present case.

First, we describe the Legendre transformation for a theory described by the action $\cS^{(m,n)}[\hat{\mathbb{C}},\check{\mathbb{C}}]$. To this end, we introduce the parent action
\begin{align}
	\label{ComplexLegendre}
	\cS^{(m,n)}[\hat{\mathbb{C}},\check{\mathbb{C}},\hat{\mathfrak{C}},\check{\mathfrak{C}}] &= \cS^{(m,n)}[\hat{\mathbb{C}},\check{\mathbb{C}}] +
	\Big \{
	\ri^{m+n+1} \int \rd^4 x \, e \, \Big(
	\hat{\mathbb{C}}^{[\D] \a(m+n)} \check{\mathfrak{C}}^{[\D]}_{\a(m+n)} \non \\
	& \qquad \qquad \qquad \qquad \qquad \qquad \qquad \qquad  + \hat{\mathfrak{C}}^{[\D] \a(m+n)} \check{\mathbb{C}}^{[\D]}_{\a(m+n)} \Big)+ \text{c.c.} 
	\Big \} ~,
\end{align}
where $\hat{\mathbb{C}}^{[\D]}_{\a(m+n)}$ and $\check{\mathbb{C}}^{[\D]}_{\a(m+n)}$ are unconstrained fields, while $\hat{\mathfrak{C}}^{[\D]}_{\a(m+n)}$ and $\check{\mathfrak{C}}^{[\D]}_{\a(m+n)}$ take the form:
\begin{align}
	\hat{\mathfrak{C}}^{[\D]}_{\a(m+n)} = (\nabla_\a{}^\bd)^n \phi^{(\rm D)}_{\a(m) \bd(n)} ~, \qquad
	\check{\mathfrak{C}}^{[\D]}_{\a(m+n)} = (\nabla_{\a}{}^{\bd})^m \bar{\phi}^{(\rm D)}_{\a(n) \bd(m)} ~,
\end{align}
where $\phi^{(\rm D)}_{\a(m) \ad(n)}$ is a Lagrange multiplier field. Now, if one varies eq. \eqref{ComplexLegendre} with respect to $\phi^{(\rm D)}_{\a(m) \ad(n)}$, the resulting equation of motion is exactly the Bianchi identity \eqref{ComplexBI}, whose general solution is given by \eqref{ComplexFS}. As a result, we recover the original self-dual model.

Next, we vary the parent action \eqref{ComplexLegendre} with respect to the unconstrained fields $\hat{\mathbb{C}}^{[\D]}_{\a(m+n)}$ and $\check{\mathbb{C}}^{[\D]}_{\a(m+n)}$. The resulting equations of motion are
\begin{align}
	\hat{\mathbb{M}}^{[\D]}_{\a(m+n)} = - \hat{\mathfrak{C}}^{[\D]}_{\a(m+n)}~, \qquad \check{\mathbb{M}}^{[\D]}_{\a(m+n)} = - \check{\mathfrak{C}}^{[\D]}_{\a(m+n)}~,
\end{align}
which we may solve to obtain $\hat{\mathbb{C}}^{[\D]}_{\a(m+n)} = \hat{\mathbb{C}}^{[\D]}_{\a(m+n)}(\hat{\mathfrak{C}},\check{\mathfrak{C}})$ and $\check{\mathbb{C}}^{[\D]}_{\a(m+n)} = \check{\mathbb{C}}^{[\D]}_{\a(m+n)}(\hat{\mathfrak{C}},\check{\mathfrak{C}})$. Inserting this solution into \eqref{ComplexLegendre}, we obtain the dual action
\begin{align}
	\label{ComplexDualAction}
	\cS^{(m,n)}_{\mathrm{Dual}}[\hat{\mathfrak{C}},\check{\mathfrak{C}}]
	:= \cS^{(m,n)}[\hat{\mathbb{C}},\check{\mathbb{C}},\hat{\mathfrak{C}},\check{\mathfrak{C}}] \Big |_{
		\substack{ \hat{\mathbb{C}} = \hat{\mathbb{C}}(\hat{\mathfrak{C}},\check{\mathfrak{C}}) \\
			\check{\mathbb{C}} = \check{\mathbb{C}}(\hat{\mathfrak{C}},\check{\mathfrak{C}})}} ~.
\end{align}

Now, assuming that the action $\cS^{(m,n)}[\hat{\mathbb{C}},\check{\mathbb{C}}]$ satisfies the self-duality equation \eqref{ComplexSDE}, we will show that it coincides with the dual action \eqref{ComplexDualAction}
\begin{align}
	\label{SDLegendre}
	\cS_{\mathrm{Dual}}^{(m,n)}[\hat{\mathbb{C}},\check{\mathbb{C}}] = \cS^{(m,n)}[\hat{\mathbb{C}},\check{\mathbb{C}}]~.
\end{align}
A routine calculation allows one to show that the following functional is invariant under infinitesimal $\sU(1)$ rotations \eqref{ComplexU(1)}
\begin{align}
	\label{ComplexInvariant}
	\cS^{(m,n)}[\hat{\mathbb{C}},\check{\mathbb{C}}] +
	\Big \{
	\frac{\ri^{m+n+1}}{2} \int \rd^4 x \, e \, \big(
	\hat{\mathbb{M}}^{[\D] \a(m+n)} \check{\mathbb{C}}_{\a(m+n)}^{[\D]}
	+ \check{\mathbb{M}}^{[\D] \a(m+n)} \hat{\mathbb{C}}^{[\D]}_{\a(m+n)} \big)+ \text{c.c.} 
	\Big \} ~.
\end{align}
Hence, it must also be invariant under the following finite duality transformations:
\begin{subequations}
	\begin{align}
		\hat{\mathbb{C}}_{\a(m+n)}^{'[\D]} &= \text{cos} \l \, \hat{\mathbb{C}}_{\a(m+n)}^{[\D]} + \text{sin} \l \, \hat{\mathbb{M}}_{\a(m+n)}^{[\D]} ~, \\
		\hat{\mathbb{M}}_{\a(m+n)}^{'[\D]} &= - \text{sin} \l \, \hat{\mathbb{C}}_{\a(m+n)}^{[\D]} + \text{cos} \l \, \hat{\mathbb{M}}_{\a(m+n)}^{[\D]}  ~, \\
		\check{\mathbb{C}}_{\a(m+n)}^{'[\D]} &= \text{cos} \l \, \check{\mathbb{C}}_{\a(m+n)}^{[\D]} + \text{sin} \l \, \check{\mathbb{M}}_{\a(m+n)}^{[\D]} ~, \\
		\check{\mathbb{M}}_{\a(m+n)}^{'[\D]} &= - \text{sin} \l \, \check{\mathbb{C}}_{\a(m+n)}^{[\D]} + \text{cos} \l \, \check{\mathbb{M}}_{\a(m+n)}^{[\D]}  ~.
	\end{align}
\end{subequations}
Performing this transformation on functional \eqref{ComplexInvariant} with $\l = \frac \pi 2$ yields
\begin{align}
	\cS^{(m,n)}[\hat{\mathbb{C}},\check{\mathbb{C}}] = 
	\cS^{(m,n)}[\hat{\mathfrak{C}},\check{\mathfrak{C}}] -
	\Big \{
	\ri^{m+n+1} \int \rd^4 x \, e \, \Big(
	\hat{\mathbb{C}}^{[\D] \a(m+n)} \check{\mathfrak{C}}^{[\D]}_{\a(m+n)}
	+ \hat{\mathfrak{C}}^{[\D] \a(m+n)} \check{\mathbb{C}}_{\a(m+n)}^{[\D]} \Big) + \text{c.c.} 
	\Big \} ~, 
\end{align}
and upon inserting this expression into \eqref{ComplexDualAction}, we obtain \eqref{SDLegendre}. Thus, the Lagrangian associated with the self-dual theory is invariant under Legendre transformations.

\section{Superconformal gauge multiplets}
\label{Section3}

In this section we extend the formalism of duality rotations for conformal gauge fields reviewed above to the case of $\cN$-extended superconformal gauge multiplets in conformally flat backgrounds. We recall that the latter are described by complex tensor superfields $\U_{\a(m)\ad(n)}$, $m,n \geq 0$. They are defined modulo the gauge transformations \cite{KR21,ERThesis}:
\begin{subequations}
\label{SCGT}
\begin{align}
m,n \geq 1 :& \qquad \d_\z \U_{\a(m)\ad(n)} = \nabla_\a^i {\z}_{\a(m-1) \ad(n) i} - \bar{\nabla}_{\ad i} \z_{\a(m) \ad(n-1)}{}^i  ~, \\
m \geq 1, n = 0 :& \qquad \d_\z \U_{\a(m)} = \nabla_\a^i \z_{\a(m-1) i} + \bar{\nabla}_{ij} \z_{\a(m)}{}^{ij}  ~, \\
m=n=0 :& \qquad \d_\z \U = \nabla^{ij} \z_{ij} + \bar{\nabla}_{ij} \bar{\z}^{ij}  ~, \label{SCGTc}
\end{align}
where $\nabla_A = (\nabla_a,\nabla_\a^i,\bar{\nabla}^\ad_i)$ are the covariant derivatives of $\cN$-extended conformal superspace and we have introduced the second-order operators
\begin{align}
	\nabla^{ij} = \nabla^{\a (i} \nabla_\a^{j)} ~, \qquad \bar{\nabla}_{ij} = \bar{\nabla}_{\ad (i} \bar{\nabla}_{j)}^\ad ~.
\end{align}
\end{subequations}
It should be pointed out that, for $\cN=2$, the transformation law \eqref{SCGTc} describes the linearised $\cN=2$ conformal supergravity multiplet \cite{KT}.
Further, for $\cN=1$, transformations \eqref{SCGT} are equivalent to those given in \cite{KMT,KP,KPR}:
\begin{subequations}
\begin{align}
	m,n \geq 1 :& \qquad \d_\z \U_{\a(m)\ad(n)} = \nabla_\a {\z}_{\a(m-1) \ad(n)} - \bar{\nabla}_{\ad} \z_{\a(m) \ad(n-1)} ~, \\
	m \geq 1, n = 0 :& \qquad \d_\z \U_{\a(m)} = \nabla_\a \z_{\a(m-1)} + \z_{\a(m)}  ~, \qquad \bar{\nabla}_\ad \z_{\a(m)} = 0~, \\
	m=n=0 :& \qquad \d_\z \U = \z + \bar{\z}  ~, \qquad \bar{\nabla}_\ad \z = 0~.
\end{align}
\end{subequations}

The requirement that \eqref{SCGT} is consistent with superconformal symmetry implies that $\U_{\a(m) \ad(n)}$ is primary, $K^B \U_{\a(m) \ad(n)}=0$, and its dilatation weight and $\sU(1)_R$ charge are as follows\footnote{We emphasise that, for $\cN=4$, the latter condition should be omitted.}
\begin{align}
	\mathbb{D} \U_{\a(m) \ad(n)} = -\hf \Big(m+n+4\cN-4\Big) \U_{\a(m) \ad(n)} ~, \quad \mathbb{Y} \U_{\a(m) \ad(n)} = -\frac{\cN(m-n)}{\cN-4} \U_{\a(m) \ad(n)} ~,
\end{align}
where $K^A = (K^a,S^\a_i,\bar{S}_\ad^i)$ denotes the special superconformal and $\sU(1)_R$ generators, respectively.
We note that, if $m=n=s$, the gauge prepotential $\U_{\a(s) \ad(s)}$ may be consistently restricted to be real; $\overline{\U_{\a(s)\ad(s)}} = \U_{\a(s) \ad(s)}$. In this case, the gauge transformations \eqref{SCGT} reduce to:
\begin{subequations}
	\begin{align}
		s\geq1 :& \qquad \d_\z \U_{\a(s)\ad(s)} = \nabla_\a^i \bar{\z}_{\a(s-1) \ad(s) i} - \bar{\nabla}_{\ad i} \z_{\a(s) \ad(s-1)}{}^i  ~, \label{HSTGT}\\
		s=0 :& \qquad \d_\z \U = \nabla^{ij} \bar{\z}_{ij} + \bar{\nabla}_{ij} {\z}^{ij}  ~.
	\end{align}
\end{subequations}
It should be pointed out that the flat-superspace version of eq. \eqref{HSTGT} first appeared in \cite{HST}. For $\cN\leq 2$, $\sU(1)$ duality-invariant models for $\U_{\a(s) \ad(s)}$ were described in \cite{KR21-2}.

From $\U_{\a(m) \ad(n)}$, and its conjugate $\bar{\U}_{\a(n) \ad(m)}$, we may construct the chiral field strengths
\begin{subequations}
\label{SUSYFS}
\begin{align}
	\hat{\mathbb{W}}_{\a(m+n+\cN)}^{[\D]} &= \bar{\nabla}^{2\cN} (\nabla_\a{}^\bd)^n \nabla_{\a(\cN)} \U_{\a(m) \bd(n)} ~, \qquad \bar{\nabla}^\ad_i \hat{\mathbb{W}}_{\a(m+n+\cN)}^{[\D]} = 0~, \\
	\check{\mathbb{W}}^{[\D]}_{\a(m+n+\cN)} &= \bar{\nabla}^{2\cN} (\nabla_{\a}{}^{\bd})^m \nabla_{\a(\cN)} \bar{\U}_{\a(n) \bd(m)} ~, \qquad \bar{\nabla}^\ad_i \check{\mathbb{W}}^{[\D]}_{\a(m+n+\cN)} = 0~,
\end{align}
\end{subequations}
where we recall that $\D = m-n$, we have made the definitions:
\begin{align}
	\nabla_{\a(\cN)} = \ve_{i_1 \dots i_\cN} \nabla_{(\a_1}^{i_1} \dots \nabla_{\a_\cN)}^{i_\cN} ~, \quad \bar{\nabla}^{\ad(\cN)} = \ve^{i_1 \dots i_\cN} \bar{\nabla}^{(\ad_1}_{i_1} \dots \bar{\nabla}^{\ad_\cN)}_{i_\cN}~, \quad
	\bar{\nabla}^{2 \cN}= \bar{\nabla}_{\ad(\cN)} \bar{\nabla}^{\ad(\cN)} ~,
\end{align}
and the totally antisymmetric tensor $\ve_{i_1 \dots i_\cN}$ is normalised as $\ve^{1 \dots \cN} = \ve_{1 \dots \cN}  = 1$. It should be emphasised that, if $m=n=s$, these descendants coincide; $\hat{\mathbb{W}}^{[0]}_{\a(2s+\cN)} = \check{\mathbb{W}}^{[0]}_{\a(2s+\cN)}$~. Their superconformal transformation laws are characterised by the properties:
\begin{subequations} 
	\bea 
	K^B \hat{\mathbb{W}}^{[\D]}_{\a(m+n+\cN)} &=&0~, \qquad 
	\mathbb{D} \hat{\mathbb{W}}^{[\D]}_{\a(m+n+\cN)} = \hf\Big(4 - \D -\cN\Big) \hat{\mathbb{W}}^{[\D]}_{\a(m+n+\cN)}~,\\
	K^{B} \check{\mathbb{W}}^{[\D]}_{\a(m+n+\cN)}&=&0~,\qquad \mathbb{D}\check{{\mathbb{W}}}^{[\D]}_{\a(m+n+\cN)}
	=\hf\Big(4 + \D -\cN\Big)\check{{\mathbb{W}}}^{[\D]}_{\a(m+n+\cN)}~.
	\eea
\end{subequations} 
Further, on conformally flat backgrounds, these descendants are gauge-invariant
\bea
\d_\z \hat{\mathbb{W}}^{[\D]}_{\a(m+n+\cN)} = \d_\z  \check{\mathbb{W}}^{[\D]}_{\a(m+n+\cN)} =0~.
\eea
In what follows, we will assume such a geometry. It is important to note that, in such backgrounds, the field strengths \eqref{ComplexFS} obey the following Bianchi identity
\be
\label{SCHSBI}
(\nabla^\b{}_\ad)^m \nabla^{\b(\cN)} \hat{\mathbb{W}}^{[\D]}_{\a(n) \b(m+\cN)} 
= (-1)^{\cN(m+n+1)} (\nabla_\a{}^\bd)^n \bar{\nabla}^{\bd(\cN)}\bar{\check{\mathbb{W}}}^{[\D]}_{\ad(m) \bd(n+\cN)} ~.
\ee

\subsection{$\sU(1)$ duality rotations for superconformal gauge superfields}

Considering $\cS^{(m,n;\cN)}[\hat{\mathbb{W}},\check{\mathbb{W}}]$ as a functional of the chiral, but otherwise unconstrained, superfields $\hat{\mathbb{W}}^{[\D]}_{\a(m+n+\cN)}$, $\check{\mathbb{W}}^{[\D]}_{\a(m+n+\cN)}$ and their conjugates, we define the dual tensors
\begin{subequations}
\label{DualSfs}
\begin{align}
\ri^{m+n+1} \hat{\mathbb{M}}^{[\D]}_{\a(m+n+\cN)} &:=  \frac{\d \cS^{(m,n;\cN)}[\hat{\mathbb{W}},\check{\mathbb{W}}]}{\d \check{\mathbb{W}}^{[\D] \a(m+n+\cN)}} ~, \qquad \bar{\nabla}^\ad_i \hat{\mathbb{M}}^{[\D]}_{\a(m+n+\cN)} = 0~, \\
\qquad \ri^{m+n+1} \check{\mathbb{M}}^{[\D]}_{\a(m+n+\cN)} &:=  \frac{\d \cS^{(m,n;\cN)}[\hat{\mathbb{W}},\check{\mathbb{W}}]}{\d \hat{\mathbb{W}}^{[\D] \a(m+n+\cN)}}~, \qquad
\bar{\nabla}^\ad_i \check{\mathbb{M}}^{[\D]}_{\a(m+n+\cN)} = 0~,
\label{5.15}
\end{align}
\end{subequations}
where the variational derivative is defined as follows
\begin{align}
	\d \cS^{(m,n;\cN)}[\hat{\mathbb{W}},\check{\mathbb{W}}] &= \int \rd^{4}x \, \rd^{2\cN} \q \, {\mathcal E} \, 
	\Big \{ \d \hat{\mathbb{W}}^{[\D] \a(m+n+\cN)} \frac{\d \cS^{(m,n;\cN)}[\hat{\mathbb{W}},\check{\mathbb{W}}]}{\d \hat{\mathbb{W}}^{[\D] \a(m+n+\cN)}}\non  \\
	&\qquad\qquad\qquad\qquad\qquad
	+ \d \check{\mathbb{W}}^{[\D] \a(m+n+\cN)} \frac{\d \cS^{(m,n;\cN)}[\hat{\mathbb{W}},\check{\mathbb{W}}]}{\d \check{\mathbb{W}}^{[\D] \a(m+n+\cN)}} \Big \} + \text{c.c.}
\end{align}
Here ${\mathcal E}$ denotes the chiral measure. The superconformal transformation law of the dual superfields \eqref{DualSfs} are characterised by the properties: 
\begin{subequations} 
	\bea 
	K^B \hat{\mathbb{M}}^{[\D]}_{\a(m+n+\cN)} &=&0~, \qquad 
	\mathbb{D} \hat{\mathbb{M}}^{[\D]}_{\a(m+n+\cN)} = \hf\Big(4 - \D -\cN\Big) \hat{\mathbb{M}}^{[\D]}_{\a(m+n+\cN)}~, \\
	K^B \check{{\mathbb{M}}}^{[\D]}_{\a(m+n+\cN)}&=&0~,\qquad \mathbb{D}\check{{\mathbb{M}}}^{[\D]}_{\a(m+n+\cN)}
	=\hf\Big(4 + \D -\cN\Big) \check{{\mathbb{M}}}^{[\D]}_{\a(m+n+\cN)}~.
	\eea
\end{subequations} 
Additionally, varying $\cS^{(m,n;\cN)}[\hat{\mathbb{W}},\check{\mathbb{W}}]$ with respect to $\bar{\U}^{\a(n) \ad(m)}$ yields the following equation of motion
\bea
\label{SCHSEoM}
(\nabla^\b{}_\ad)^m \nabla^{\b(\cN)} \hat{\mathbb{M}}^{[\D]}_{\a(n) \b(m+\cN)} 
= (-1)^{\cN(m+n+1)} (\nabla_\a{}^\bd)^n \bar{\nabla}^{\bd(\cN)}\bar{\check{\mathbb{M}}}^{[\D]}_{\ad(m) \bd(n+\cN)} ~.
\eea

It is clear to see that the Bianchi identity \eqref{SCHSBI} and the equation of motion \eqref{SCHSEoM} are together invariant under the following $\sU(1)$ duality rotations:
\begin{subequations}
	\label{SCComplexU(1)}
	\begin{align}
		\d_\l \hat{\mathbb{W}}^{[\D]}_{\a(m+n+\cN)} &= \l \hat{\mathbb{M}}^{[\D]}_{\a(m+n+\cN)} ~, \quad \d_\l \hat{\mathbb{M}}^{[\D]}_{\a(m+n+\cN)} = - \l \hat{\mathbb{W}}^{[\D]}_{\a(m+n+\cN)} ~, \\
		\d_\l \check{\mathbb{W}}^{[\D]}_{\a(m+n+\cN)} &= \l \check{\mathbb{M}}^{[\D]}_{\a(m+n+\cN)} ~, \quad \d_\l \check{\mathbb{M}}^{[\D]}_{\a(m+n+\cN)} = - \l \check{\mathbb{W}}^{[\D]}_{\a(m+n+\cN)} ~.
	\end{align}
\end{subequations}
Here $\l = \bar{\l}$ is an arbitrary constant parameter. A routine analysis then leads to the self-duality equation for $\cS^{(m,n;\cN)}[\hat{\mathbb{W}},\check{\mathbb{W}}]$
\bea
\label{SDeqSUSY}
\ri^{m+n+1} \int \rd^{4}x \, \rd^{2\cN} \q \, {\mathcal E} \, \Big \{ \hat{\mathbb{W}}^{[\D] \a(m+n+\cN)}  \check{\mathbb{W}}^{[\D]}_{\a(m+n+\cN)}
+ \hat{\mathbb{M}}^{[\D] \a(m+n+\cN)} \check{\mathbb{M}}^{[\D]}_{\a(m+n+\cN)} \Big \} + \text{c.c.}  = 0 ~.~~~~~~~
\eea
We emphasise that this equation must hold for chiral, but otherwise unconstrained, superfields $\hat{\mathbb{W}}^{[\D]}_{\a(m+n+\cN)}$ and $\check{\mathbb{W}}^{[\D]}_{\a(m+n+\cN)}$.

The simplest solution of the self-duality equation \eqref{SDeqSUSY} is the free action
\bea
\label{FreeAction}
\cS^{(m,n;\cN)}_{\rm Free}[\hat{\mathbb{W}},\check{\mathbb{W}}]
= {\ri^{m+n}}\int \rd^{4}x \, \rd^{2\cN} \q \, {\mathcal E} \,  \hat{ {\mathbb{W}}}^{[\D] \a(m+n+\cN)}\check{{\mathbb{W}}}^{[\D]}_{\a(m+n+\cN)} 
+{\rm c.c.}
\eea
For $\cN=1$, the flat-superspace version of \eqref{FreeAction} was first given in \cite{KMT} and then extended to general conformally flat backgrounds in \cite{KP}. The extension to $\cN > 1$ followed shortly thereafter \cite{KR21}.  

\subsection{Self-duality under Legendre transformations}

In section \ref{section2.2}, we extended the well-known result that $\sU(1)$ duality invariance implies self-duality under Legendre transformations to the case of general conformal gauge fields. Here, we will extend this result to the case of a superconformal gauge multiplet.

We begin by introducing the parent action
\begin{align}
	\label{SCComplexLegendre}
	\cS^{(m,n;\cN)}[\hat{\mathbb{W}},\check{\mathbb{W}},\hat{\mathfrak{W}},\check{\mathfrak{W}}] &= \cS^{(m,n;\cN)}[\hat{\mathbb{W}},\check{\mathbb{W}}] +
	\Big \{
	\ri^{m+n+1} \int \rd^4 x \, \rd^{2\cN} \q \, \mathcal{E} \, \big(
	\hat{\mathbb{W}}^{[\D] \a(m+n+\cN)} \check{\mathfrak{W}}^{[\D]}_{\a(m+n+\cN)} \non \\
	& \qquad \qquad \qquad \qquad \qquad \quad + \hat{\mathfrak{W}}^{[\D] \a(m+n+\cN)} \check{\mathbb{W}}^{[\D]}_{\a(m+n+\cN)} \big)+ \text{c.c.} 
	\Big \} ~,
\end{align}
where $\hat{\mathbb{W}}^{[\D]}_{\a(m+n+\cN)}$ and $\check{\mathbb{W}}^{[\D]}_{\a(m+n+\cN)}$ are chiral, but otherwise unconstrained, superfields, while $\hat{\mathfrak{W}}^{[\D]}_{\a(m+n+\cN)}$ and $\check{\mathfrak{W}}^{[\D]}_{\a(m+n+\cN)}$ take the form:
\begin{subequations}
\begin{align}
	\hat{\mathfrak{W}}^{[\D]}_{\a(m+n+\cN)} &= \bar{\nabla}^{2\cN} (\nabla_\a{}^\bd)^n \nabla_{\a(\cN)} \U^{(\rm D)}_{\a(m) \bd(n)} ~, \\
	\check{\mathfrak{W}}^{[\D]}_{\a(m+n+\cN)} &= \bar{\nabla}^{2\cN} (\nabla_{\a}{}^{\bd})^m \nabla_{\a(\cN)} \bar{\U}^{(\rm D)}_{\a(n) \bd(m)} ~.
\end{align}
\end{subequations}
Here $\U^{(\rm D)}_{\a(m) \ad(n)}$ is a Lagrange multiplier superfield; if one varies eq. \eqref{SCComplexLegendre} with respect to $\U^{(\rm D)}_{\a(m) \ad(n)}$, the resulting equation of motion is exactly the Bianchi identity \eqref{SCHSBI}, whose general solution is given by \eqref{SUSYFS}. Consequently, we recover the original model.

Next, varying \eqref{SCComplexLegendre} with respect to $\hat{\mathbb{W}}^{[\D]}_{\a(m+n+\cN)}$ and $\check{\mathbb{W}}^{[\D]}_{\a(m+n+\cN)}$ leads to
\begin{align}
	\hat{\mathbb{M}}^{[\D]}_{\a(m+n+\cN)} = - \hat{\mathfrak{W}}^{[\D]}_{\a(m+n+\cN)}~, \qquad \check{\mathbb{W}}^{[\D]}_{\a(m+n+\cN)} = - \check{\mathfrak{W}}^{[\D]}_{\a(m+n+\cN)}~,
\end{align}
which we may solve to obtain 
\bea
\hat{\mathbb{W}}^{[\D]}_{\a(m+n+\cN)} = \hat{\mathbb{W}}^{[\D]}_{\a(m+n+\cN)}(\hat{\mathfrak{W}},\check{\mathfrak{W}})~, \qquad 
\check{\mathbb{W}}^{[\D]}_{\a(m+n+\cN)} = \check{\mathbb{W}}^{[\D]}_{\a(m+n+\cN)}(\hat{\mathfrak{W}},\check{\mathfrak{W}})~. 
\eea
Inserting this solution into \eqref{SCComplexLegendre} leads to the dual action
\begin{align}
	\label{SCComplexDualAction}
	\cS^{(m,n;\cN)}_{\mathrm{Dual}}[\hat{\mathfrak{W}},\check{\mathfrak{W}}]
	:= \cS^{(m,n;\cN)}[\hat{\mathbb{W}},\check{\mathbb{W}},\hat{\mathfrak{W}},\check{\mathfrak{W}}] \Big |_{
		\substack{ \hat{\mathbb{W}} = \hat{\mathbb{W}}(\hat{\mathfrak{W}},\check{\mathfrak{W}}) \\
			\check{\mathbb{W}} = \check{\mathbb{W}}(\hat{\mathfrak{W}},\check{\mathfrak{W}})}} ~.
\end{align}

Now, assuming that the action $\cS^{(m,n;\cN)}[\hat{\mathbb{W}},\check{\mathbb{W}}]$ satisfies the self-duality equation \eqref{SDeqSUSY}, we will show that it coincides with the dual action \eqref{SCComplexDualAction}
\begin{align}
	\label{SCSDLegendre}
	\cS^{(m,n;\cN)}_{\mathrm{Dual}}[\hat{\mathbb{W}},\check{\mathbb{W}}] = \cS^{(m,n;\cN)}[\hat{\mathbb{W}},\check{\mathbb{W}}]~.
\end{align}
A routine calculation allows one to show that the following functional is invariant under infinitesimal $\sU(1)$ rotations \eqref{SCComplexU(1)}
\begin{align}
	\label{SCComplexInvariant}
	\cS^{(m,n;\cN)}[\hat{\mathbb{W}},\check{\mathbb{W}}] +
	\Big \{
	\frac{\ri^{m+n+1}}{2} \int \rd^4 x \, \rd^{2\cN} \q \, {\mathcal E} \, \big(&
	\hat{\mathbb{M}}^{[\D] \a(m+n+\cN)} \check{\mathbb{W}}_{\a(m+n+\cN)}^{[\D]} \non \\
	& \qquad
	+ \check{\mathbb{M}}^{[\D] \a(m+n+\cN)} \hat{\mathbb{W}}^{[\D]}_{\a(m+n+\cN)} \big)+ \text{c.c.} 
	\Big \} ~.
\end{align}
Hence, it must also be invariant under the following finite duality transformations:
\begin{subequations}
	\begin{align}
		\hat{\mathbb{W}}_{\a(m+n+\cN)}^{'[\D]} &= \text{cos} \l \, \hat{\mathbb{W}}_{\a(m+n+\cN)}^{[\D]} + \text{sin} \l \, \hat{\mathbb{M}}_{\a(m+n+\cN)}^{[\D]} ~, \\
		\hat{\mathbb{M}}_{\a(m+n+\cN)}^{'[\D]} &= - \text{sin} \l \, \hat{\mathbb{W}}_{\a(m+n+\cN)}^{[\D]} + \text{cos} \l \, \hat{\mathbb{M}}_{\a(m+n+\cN)}^{[\D]}  ~, \\
		\check{\mathbb{W}}_{\a(m+n+\cN)}^{'[\D]} &= \text{cos} \l \, \check{\mathbb{W}}_{\a(m+n+\cN)}^{[\D]} + \text{sin} \l \, \check{\mathbb{M}}_{\a(m+n+\cN)}^{[\D]} ~, \\
		\check{\mathbb{M}}_{\a(m+n+\cN)}^{'[\D]} &= - \text{sin} \l \, \check{\mathbb{W}}_{\a(m+n+\cN)}^{[\D]} + \text{cos} \l \, \check{\mathbb{M}}_{\a(m+n+\cN)}^{[\D]}  ~.
	\end{align}
\end{subequations}
Performing this transformation on \eqref{SCComplexInvariant} with $\l = \frac \pi 2$ yields
\begin{align}
	\cS^{(m,n;\cN)}[\hat{\mathbb{W}},\check{\mathbb{W}}] &= 
	\cS^{(m,n;\cN)}[\hat{\mathfrak{W}},\check{\mathfrak{W}}] -
	\Big \{
	\ri^{m+n+1} \int \rd^4 x \, \rd^{2\cN} \q \, {\mathcal E} \, \Big(
	\hat{\mathbb{W}}^{[\D] \a(m+n+\cN)} \check{\mathfrak{W}}^{[\D]}_{\a(m+n+\cN)} \non \\
	& \qquad \qquad \qquad \qquad \qquad \qquad 
	+ \hat{\mathfrak{W}}^{[\D] \a(m+n+\cN)} \check{\mathbb{W}}_{\a(m+n+\cN)}^{[\D]} \Big) + \text{c.c.} 
	\Big \} ~, 
\end{align}
and upon inserting this expression into \eqref{SCComplexDualAction}, we obtain \eqref{SCSDLegendre}.

\section{The $\cN=2$ superconformal gravitino multiplet}
\label{Section4}

Consider a dynamical system describing the propagation of the $\cN=2$ superconformal gravitino multiplet in curved superspace, see e.g. \cite{KRTM2} for a review of the latter. The assoicated action functional $\cS[\hat{\mathbb{W}},\check{\mathbb{W}}]$ is assumed to depend on the chiral field strengths $\hat{\mathbb{W}}_\a$, $\check{\mathbb{W}}_\a$ and their conjugates, which are defined as follows:
\begin{align}
	\label{2.1}
	\hat{\mathbb{W}}_\a = \bar{\nabla}^{4} \nabla_\a^i {\U}_i~, \qquad \check{\mathbb{W}}_\a = \bar{\nabla}^{4} \nabla^{ij} \nabla_{\a i} {\bar \U}_j ~.
\end{align}
Here $\nabla_A = (\nabla_a,\nabla_\a^i, \bar{\nabla}^\ad_i)$ denote the $\cN=2$ conformally covariant derivatives and the gauge prepotential $\U_i$ is defined modulo gauge transformations of the form
\begin{align}
	\label{2.3}
	\d_{\z} \U_i = {\nabla}^{jk} \z_{ijk} + \bar{\nabla}^{\ad j} \z_{\ad ij}~.
\end{align}
Gauge transformation \eqref{2.3} is superconformal provided $\U_i$ is characterised by the properties
\begin{align}
	\label{2.4}
	K^B \U_i = 0 ~, \qquad \mathbb{D} \U_i = - \mathbb{Y} \U_i = -2 \U_i~,
\end{align}
which imply that the field strengths are primary in general curved backgrounds:
\begin{subequations}
	\label{2.5}
	\begin{align}
		K^B \hat{\mathbb{W}}_\a  &= 0 ~, \qquad \mathbb{D} \hat{\mathbb{W}}_\a  = \frac{1}{2} \hat{\mathbb{W}}_\a  ~, \qquad \mathbb{Y} \hat{\mathbb{W}}_\a  = - \hat{\mathbb{W}}_\a ~, \\
		K^B \check{\mathbb{W}}_\a &= 0 ~, \qquad \mathbb{D} \check{\mathbb{W}}_\a = \frac{3}{2} \check{\mathbb{W}}_\a ~, \qquad \mathbb{Y} \check{\mathbb{W}}_\a = - 3 \check{\mathbb{W}}_\a ~.
	\end{align}
\end{subequations}
However, the gauge transformation \eqref{2.3} leaves the field strengths \eqref{2.1} invariant only in conformally flat backgrounds
\be
\label{2.6}
W_{\a \b} = 0 \quad \implies \quad \d_{\z} \hat{\mathbb{W}}_{\a} = \d_{\z} \check{\mathbb{W}}_{\a} = 0~.
\ee
Here $W_{\a \b}$ denotes the $\cN=2$ super-Weyl tensor. Such a geometry will be assumed in what follows. It is important to note that the field strengths \eqref{2.1} satisfy the Bianchi identity
\begin{align}
	\label{2.7}
	\nabla^\a_j \nabla^{ij} \hat{\mathbb W}_\a = \bar{\nabla}^{\ad i} \bar{\check{\mathbb{W}}}_\ad~.
\end{align}

\subsection{$\sU(1)$ duality-invariant models}
\label{section2.1}

We now consider $\cS[\hat{\mathbb{W}},\check{\mathbb{W}}]$ as a functional of the unconstrained fields $\hat{\mathbb{W}}_{\a}$, $\check{\mathbb{W}}_{\a}$ and their conjugates. This allows us to introduce the dual chiral superfields
\begin{subequations}
\label{2.8}
\begin{align}
\hat{\mathbb{M}}_{\a} &:= \ri  \frac{\d \cS[\hat{\mathbb{W}},\check{\mathbb{W}}]}{\d \check{\mathbb{W}}^{\a}} ~, \qquad \bar{\nabla}^\ad_i \hat{\mathbb{M}}_{\a} = 0 ~, \\
\quad \check{\mathbb{M}}_{\a} &:= \ri \frac{\d \cS[\hat{\mathbb{W}},\check{\mathbb{W}}]}{\d \hat{\mathbb{W}}^{\a}}~, \qquad \bar{\nabla}^\ad_i \check{\mathbb{M}}_{\a} = 0 ~,
\end{align}
\end{subequations}
where we have made the definition
\begin{align}
	\label{2.9}
	\d \cS[\hat{\mathbb{W}},\check{\mathbb{W}}] = \int \rd^4x \rd^4\q \, \cE \, 
	\Bigg \{ \d \hat{\mathbb{W}}^{\a} \frac{\d \cS[\hat{\mathbb{W}},\check{\mathbb{W}}]}{\d \hat{\mathbb{W}}^{\a}} + \d \check{\mathbb{W}}^{\a} \frac{\d \cS[\hat{\mathbb{W}},\check{\mathbb{W}}]}{\d \check{\mathbb{W}}^{\a}} \Bigg \} + \text{c.c.}
\end{align}
The superconformal properties of the superfields \eqref{2.8} are: 
\begin{subequations} 
	\label{2.10}
	\begin{align}
		K^B \hat{\mathbb{M}}_\a  &= 0 ~, \qquad \mathbb{D} \hat{\mathbb{M}}_\a  = \frac{1}{2} \hat{\mathbb{M}}_\a  ~, \qquad \mathbb{Y} \hat{\mathbb{M}}_\a  = - \hat{\mathbb{M}}_\a ~, \\
		K^B \check{\mathbb{M}}_\a &= 0 ~, \qquad \mathbb{D} \check{\mathbb{M}}_\a = \frac{3}{2} \check{\mathbb{M}}_\a ~, \qquad \mathbb{Y} \check{\mathbb{M}}_\a = - 3 \check{\mathbb{M}}_\a ~.
	\end{align}
\end{subequations} 
Varying $\cS[\hat{\mathbb{W}},\check{\mathbb{W}}]$ with respect to $\U_i$ yields the equation of motion
\begin{align}
	\label{2.11}
	\nabla^\a_j \nabla^{ij} \hat{\mathbb M}_\a = \bar{\nabla}^{\ad i} \bar{\check{\mathbb{M}}}_\ad~.
\end{align}
whose functional form mirrors that of the Bianchi identity \eqref{2.7}.

It is clear from the discussion above that the system of equations \eqref{2.7} and \eqref{2.11} is invariant under the $\sU(1)$ duality rotations
\begin{subequations}
	\label{2.12}
	\begin{align}
		\d_\l \hat{\mathbb{W}}_{\a} &= \l \hat{\mathbb{M}}_{\a} ~, \quad  \d_\l \hat{\mathbb{M}}_{\a} = - \l \hat{\mathbb{W}}_{\a} ~, \\
		\d_\l \check{\mathbb{W}}_{\a} &= \l \check{\mathbb{M}}_{\a}  ~, \quad \d_\l \check{\mathbb{M}}_{\a} = - \l \check{\mathbb{W}}_{\a} ~.
	\end{align}
\end{subequations}
One may construct $\sU(1)$ duality-invariant models for $\U_i$.
Their actions satisfy  the self-duality equation
\bea
\label{2.13}
\mathrm{Im} \int \rd^4 x \rd^4\q \, \cE \, \Big \{ \hat{\mathbb{W}}^{\a}  \check{\mathbb{W}}_{\a}
+ \hat{\mathbb{M}}^{\a} \check{\mathbb{M}}_{\a} \Big \}  = 0 ~,
\eea
which must hold for unconstrained fields $\hat{\mathbb{W}}_{\a}$ and $\check{\mathbb{W}}_{\a}$.
The simplest solution of this equation is the free action\footnote{The $\cN=1$ superspace reduction of this action is described in appendix \ref{appendixB}.}
\bea
\label{2.14}
\cS_{\mathrm{Free}}[\hat{\mathbb{W}}, \check{\mathbb{W}}] = \int \rd^4 x \rd^4\q \, \cE \, \hat{\mathbb{W}}^{\a} \check{\mathbb{W}}_{\a} + \text{c.c.}
\eea

\subsection{Self-duality under Legendre transformations} \label{section4.2}

We begin by describing a Legendre transformation for a generic theory with action 
$\cS[\hat{\mathbb{W}},\check{\mathbb{W}}]$.
For this we introduce the parent action
\begin{align}
	\label{2.15}
	\cS[\hat{\mathbb{W}},\check{\mathbb{W}},\hat{\mathfrak{W}},\check{\mathfrak{W}}] = \cS[\hat{\mathbb{W}},\check{\mathbb{W}}] +
	 \Big \{
	  \ri \int \rd^4 x \rd^4 \q \, \cE \, \big(
	  \hat{\mathbb{W}}^\a \check{\mathfrak{W}}_\a 
	  + \hat{\mathfrak{W}}^{\a} \check{\mathbb{W}}_\a \big)+ \text{c.c.} 
\Big \} ~.
\end{align}
Here $\hat{\mathbb{W}}_\a$ and $\check{\mathbb{W}}_\a$ are chiral, but otherwise unconstrained superfields, while $\hat{\mathfrak{W}}_\a$ and $\check{\mathfrak{W}}_\a$ take the form
\begin{align}
	\label{2.16}
	 \check{\mathfrak{W}}_\a := \bar{\nabla}^{4} \nabla_\a^i {\U}_i^{\rm D} ~, \qquad \hat{\mathfrak{W}}_\a := \bar{\nabla}^{4} \nabla^{ij} \nabla_{\a i} \bar{\U}_j^{\rm D} ~,
\end{align}
where $\U_i^{\rm D}$ is a Lagrange multiplier superfield. Indeed, upon varying \eqref{2.15} with respect to 
$\U_i^{\rm D}$ one obtains the Bianchi identity \eqref{2.7}, and its general solution is given by eq. \eqref{2.1}, for some primary isospinor superfield $\U_i$ defined modulo the gauge transformations \eqref{2.3} and characterised by the superconformal properties \eqref{2.4}. 
As a result the second term in \eqref{2.15} becomes a total derivative, and we end up with the original model. Alternatively, if we first vary the parent action with respect to $\hat{\mathbb{W}}_\a$ and $\check{\mathbb{W}}_\a$, the equations of motion are
\begin{align}
	\label{2.17}
	\hat{\mathbb{M}}_{\a} = - \hat{\mathfrak{W}}_{\a}~, \qquad \check{\mathbb{M}}_{\a} = - \check{\mathfrak{W}}_{\a}~,
\end{align}
which we may solve to express $\hat{\mathbb{W}}_\a$ and $\check{\mathbb{W}}_\a$ in terms of the dual field strengths.
Inserting this solution into \eqref{2.13}, we obtain the dual model
\begin{align}
	\label{2.18}
	\cS_{\mathrm{Dual}}[\hat{\mathfrak{W}},\check{\mathfrak{W}}]
	:= \cS[\hat{\mathbb{W}},\check{\mathbb{W}},\hat{\mathfrak{W}},\check{\mathfrak{W}}] \Big |_{
		\substack{ \hat{\mathbb{W}} = \hat{\mathbb{W}}(\hat{\mathfrak{W}},\check{\mathfrak{W}}) \\
			\check{\mathbb{W}} = \check{\mathbb{W}}(\hat{\mathfrak{W}},\check{\mathfrak{W}})}} ~.
\end{align}

Now, given an action $\cS[\hat{\mathbb{W}},\check{\mathbb{W}}]$ satisfying the self-duality equation \eqref{2.10}, our aim is to show that it satisfies
\begin{align}
	\label{2.19}
	\cS_{\mathrm{Dual}}[\hat{\mathbb{W}},\check{\mathbb{W}}] = \cS[\hat{\mathbb{W}},\check{\mathbb{W}}]~,
\end{align}
which means that the corresponding Lagrangian is invariant under Legendre transformations. A routine calculation allows one to show that the following functional
\begin{align}
	\label{2.20}
	\cS[\hat{\mathbb{W}},\check{\mathbb{W}}] +
	\Big \{
	\frac{\ri}{2} \int \rd^4 x \rd^4 \q \, \cE \, \big(
	\hat{\mathbb{M}}^\a \check{\mathbb{W}}_\a 
	+ \check{\mathbb{M}}^{\a} \hat{\mathbb{W}}_\a \big)+ \text{c.c.} 
	\Big \} ~.
\end{align}
is invariant under \eqref{2.9}. The latter may be exponentiated to obtain the finite $\sU(1)$ duality transformations
\begin{subequations}
	\label{2.21}
\begin{align}
	\hat{\mathbb{W}}'_\a &= \text{cos} \l \, \hat{\mathbb{W}}_\a + \text{sin} \l \, \hat{\mathbb{M}}_\a ~, \qquad
	\hat{\mathbb{M}}'_\a = - \text{sin} \l \, \hat{\mathbb{W}}_\a + \text{cos} \l \, \hat{\mathbb{M}}_\a  ~, \\
	\check{\mathbb{W}}'_\a &= \text{cos} \l \, \check{\mathbb{W}}_\a + \text{sin} \l \, \check{\mathbb{M}}_\a ~, \qquad
	\check{\mathbb{M}}'_\a = - \text{sin} \l \, \check{\mathbb{W}}_\a + \text{cos} \l \, \check{\mathbb{M}}_\a  ~.
\end{align}
\end{subequations}
Performing such a transformation with $\l = \frac \pi 2$ on \eqref{2.20} yields 
\begin{align}
	\label{2.22}
	\cS[\hat{\mathbb{W}},\check{\mathbb{W}}] =
	\cS[\hat{\mathfrak{W}},\check{\mathfrak{W}}] -
	\Big \{
	\ri \int \rd^4 x \rd^4 \q \, \cE \, \big(
	\hat{\mathbb{W}}^\a \check{\mathfrak{W}}_\a 
	+ \hat{\mathfrak{W}}^{ \a} \check{\mathfrak{W}}_\a \big) + \text{c.c.} 
	\Big \} ~.
\end{align}
Upon inserting this expression into \eqref{2.15}, we obtain \eqref{2.19}.


\subsection{Auxiliary superfield formulation} \label{section2.3}

The duality-invariant models described in section \ref{section2.1} may be reformulated by using an auxiliary superfield approach in a spirit of \cite{K13,ILZ}.
Such a formulation was recently employed in the study of duality-invariant models for the $\cN=2$ vector multiplet \cite{K21} and superconformal higher-spin multiplets \cite{KR21-2}.

The starting point for such an analysis is the following action functional
\begin{align}
	\label{2.23}
	\cS [\hat{\mathbb{W}},\check{\mathbb{W}}, \hat{\eta},\check{\eta}] &= 2 \int \rd^4x 
	\rd^4 \q \, {\mathcal{E}} \, \Big \{ \hat{\eta}^\a \check{\mathbb{W}}_\a + \check{\eta}^\a \hat{\mathbb{W}}_\a - 
	\hat{\eta}^\a \check{\eta}_\a - \frac{1}{2} \hat{\mathbb{W}}^\a \check{\mathbb{W}}_\a \Big \} + \text{c.c.} + \mathfrak{S}_{\text{Int}} [\hat{\eta} , \check{\eta}] ~.
\end{align}
Here we have introduced the auxiliary superfields $\hat{\eta}_\a$ and $\check{\eta}_\a$, which are characterised by the properties:
\begin{subequations}
	\label{2.24}
	\begin{align}
		K^B \hat{\eta}_\a  &= 0 ~, \quad \bar{\nabla}^{\bd}_j \hat{\eta}_\a  = 0 ~, \quad \mathbb{D} \hat{\eta}_\a  = \frac{1}{2} \hat{\eta}_\a  ~, \quad \mathbb{Y} \hat{\eta}_\a  = - \hat{\eta}_\a ~, \\
		K^B \check{\eta}_\a &= 0 ~, \quad \bar{\nabla}^{\bd}_j \check{\eta}_\a  = 0 ~, \quad \mathbb{D} \check{\eta}_\a = \frac{3}{2} \check{\eta}_\a ~, \quad \mathbb{Y} \check{\eta}_\a = - 3 \check{\eta}_\a ~.
	\end{align}
\end{subequations}
By construction, the self-interaction $\mathfrak{S}_{\text{Int}} [\hat{\eta} , \check{\eta}]$  contains cubic and higher powers of $\hat{\eta}_\a$, $\check{\eta}_\a$ and their conjugates. Varying \eqref{2.23} with respect to the auxiliary superfields yields 
\begin{align}
	\label{2.25}
\hat{\mathbb{W}}_\a = \hat{\eta}_\a - \hf \frac{\d \mathfrak{S}_{\text{Int}} [\hat{\eta} , \check{\eta}]}{\d \check{\eta}^\a}~, \qquad
\check{\mathbb{W}}_\a = \check{\eta}_\a - \hf \frac{\d \mathfrak{S}_{\text{Int}} [\hat{\eta} , \check{\eta}]}{\d \hat{\eta}^\a}~.
\end{align}
Employing perturbation theory, these equations allow one to express $\hat{\eta}_\a$ and $\check{\eta}_\a$ as functions of $\hat{\mathbb{W}}_\a$, $\check{\mathbb{W}}_\a$ and their conjugates. This means that \eqref{2.23} is equivalent to the theory described by
\begin{align}
	\label{2.26}
	\cS[\hat{\mathbb{W}},\check{\mathbb{W}}] &= \int \rd^4x 
	\rd^4 \q \, {\mathcal{E}}\, \hat{\mathbb{W}}^\a \check{\mathbb{W}}_\a + \text{c.c.} 
	+ \cS_{\text{Int}} [\hat{\mathbb{W}},\check{\mathbb{W}}] ~.
\end{align}

We now analyse the self duality equation \eqref{2.13}. It is easily verified that  this constraint is equivalent to
\be
\label{2.27}
\text{Im} \int \rd^4x \rd^4 \q \, {\mathcal{E}}\, \bigg \{
\hat{\eta}^{\a} \frac{\d \mathfrak{S}_{\text{Int}} 
	[\hat{\eta} , \check{\eta}]}{\d \hat{\eta}^{\a}} + 
\check{\eta}^{\a} \frac{\d \mathfrak{S}_{\text{Int}} 
	[\hat{\eta} , \check{\eta}]}{\d \check{\eta}^{\a}} \bigg \} = 0 ~.
\ee
Thus, the $\sU(1)$ duality invariance of the model \eqref{2.23} is equivalent to the requirement that $\mathfrak{S}_{\text{Int}}[\hat{\eta} , \check{\eta}]$ is invariant under rigid $\sU(1)$ phase transformations
\be
\label{2.28}
\mathfrak{S}_{\text{Int}} [\re^{\ri \varphi}\hat{\eta} , \re^{\ri \varphi}\check{\eta}] = \mathfrak{S}_{\text{Int}} 
[\hat{\eta} , \check{\eta}] ~, \quad \varphi \in \mathbb{R} ~.
\ee

If we allow action \eqref{2.23} to depend on a superconformal compensator, it is trivial to construct interactions satisfying the self-duality equation \eqref{2.28}. For instance:
\begin{align}
	\label{Interactions}
	\mathfrak{S}_{\text{Int}}[\hat{\eta} , \check{\eta}; \Xi] &= \int \rd^4x \rd^4 \q \rd^4 \bar{\q} \, E \, \bigg \{
	\mathfrak{F} \big( {v \bar{v}}, {w \bar{w}} \big) + \frac{ \hat{\eta} \check{\eta} \bar{\hat{\eta}} \bar{\check{\eta}} }{\Xi \bar \X} 
	\mathfrak{G} \big( {v \bar{v}}, {w \bar{w}} \big) + \frac{ \hat{\eta}^2 \check{\eta}^2 \bar{\hat{\eta}}^2 \bar{\check{\eta}}^2 }{(\X \bar \X)^2} 
	\mathfrak{H} \big( {v \bar{v}}, {w \bar{w}} \big)
	\bigg \} ~. ~~~
\end{align}
Here $\mathfrak{F}(x,y)$, $\mathfrak{G}(x,y)$ and $\mathfrak{H}(x,y)$ are dimensionless, real functions of two real variables, and $\Xi \neq \bar{\Xi}$ is a conformal compensator with the properties:
\begin{align}
	K^A \Xi = 0 ~, \qquad \mathbb{D} \Xi = 2 \Xi ~, \qquad \mathbb{Y} \Xi = -4\X~.
\end{align}
In \eqref{Interactions} we have introduced the primary, dimensionless and uncharged descendants
\bea
v: = \bar \Xi^{-1} \nabla^4 \Big( \frac{ \hat{{\eta}} \check{{\eta}}}{\X}\Big)~, \qquad
w: = \bar \Xi^{-2} \nabla^4 \Big( \frac{ \hat{{\eta}}^2 \check{{\eta}}^2}{\X^2}\Big) ~.
\eea
We emphasise that, unless the real functions above are chosen such that the functional \eqref{Interactions} is $\Xi$-independent, the latter describes a non-superconformal theory. Below, the superconformal case will be studied in more detail.


\subsection{Superconformal duality-invariant models}

As discussed above, the interaction \eqref{Interactions} describes a superconformal theory if 
the associated 
functions $\mathfrak{F}$, $\mathfrak{G}$ and $\mathfrak{H}$
are chosen such that \eqref{Interactions} is independent of $\Xi$.
This is the case if:
\begin{align}
	\label{4.31}
	\mathfrak{F}(x,y) = \mathfrak{G}(x,y) = 0 ~, \qquad \mathfrak{H}(x,y) = \frac{1}{\sqrt{y}} \mathfrak{h} \Big(\frac{x^2}{y} \Big)~,
\end{align}
where $\mathfrak{h}(x)$ is a real function of a real variable. 

It is well known that the requirement of conformal invariance uniquely singles out the ModMax theory \cite{BLST} 
in the family of $\sU(1)$ duality-invariant models for nonlinear electrodynamics 
without higher derivatives, eq. \eqref{NLED}. This uniqueness is no longer present if the Maxwell field (conformal spin
$s=1$) is replaced by a conformal higher-spin field $s>1$.  Even in the conformal graviton case (spin $s=2$), Ref.  \cite{KR21-2} constructed a two-parameter family of conformal $\sU(1)$ duality-invariant models.  
Furthermore, the $\cN=1$ supersymmetric ModMax theory  \cite{BLST2,K21} is a unique superconformal representative in the family of $\sU(1) $ duality-invariant models for nonlinear supersymmetric electrodynamics
proposed in \cite{KT1}  (see eq. (2.10) in \cite{KT1} or eq. (2.22) in  \cite{BLST2}). However, for more general supersymmetric models of the functional form (2.10) in  \cite{BLST2}, 
a recent paper \cite{Kuzenko:2023igt} constructed a family of $\cN=1$ superconformal vector multiplet models 
with $\sU(1)$ duality invariance. It is therefore not surprising that the superconformal $\sU(1)$ duality-invariant model  
\eqref{4.31} has nontrivial functional freedom.


\section{Concluding comments}

In this paper, we have developed the general formalism of $\sU(1)$ duality rotations for the $\cN$-extended superconformal gauge multiplets $\U_{\a(m) \ad(n)}$, $m,n \geq 0$. We recall that, associated with each such multiplet is a pair of chiral field strengths $\hat{\mathbb{W}}^{[\D]}_{\a(m+n+\cN)}$ and $\check{\mathbb{W}}^{[\D]}_{\a(m+n+\cN)}$ carrying at least $\cN$ spinor indices.\footnote{It should be pointed out that, in the super-Poincar\'e case, practically all chiral field strengths given in our paper were introduced (perhaps in a somewhat disguised form) many years ago in \cite{SG81,GGRS}.} Hence, as discussed in \cite{KR21}, for any fixed $\cN>1$, chiral field strengths carrying fewer than $\cN$ indices do not originate from this family of gauge prepotentials. The $\cN=2$ story was recently completed in \cite{HKR}, where the spinor field strengths \eqref{2.1} were shown to describe the superconformal gravitino multiplet.\footnote{We recall that the scalar field strength $\mathbb{W}$ describes the vector multiplet, see e.g. \cite{KRTM2} for a review.} The completion of this analysis for $\cN>2$ remains an open problem, which would be interesting to revisit in the future.

Building on the results of \cite{HKR}, in section \ref{Section4}, we extended the formalism of duality rotations to the case of the $\cN=2$ superconformal gravitino multiplet. This formalism was then utilised, in conjunction with an auxiliary superfield formulation developed in section \ref{section2.3}, to describe new nonlinear models \eqref{Interactions}, including a family of superconformal ones \eqref{4.31}.

As discussed in the main body of this work, every duality-invariant model arises as a solution of the so-called `self-duality equation.' We reiterate that the self-duality equation for a conformal gauge field $\phi_{\a(m) \ad(n)}$, $m,n\geq1$, is \cite{KR21-2}
\bea
\label{nonsusySDeqGeneral}
\ri^{m+n+1} \int \rd^{4}x \, e \, \Big \{ \hat{\mathbb{C}}^{[\D] \a(m+n)} \check{\mathbb{C}}^{[\D]}_{\a(m+n)}
+ \hat{\mathbb{M}}^{[\D] \a(m+n)} \check{\mathbb{M}}^{[\D]}_{\a(m+n)} \Big \} + \text{c.c.}  = 0 ~,
\eea
see section \ref{Section2} for the appropriate definitions. One of the main results of this work is a supersymmetric extension of \eqref{nonsusySDeqGeneral} in the sense that it describes $\sU(1)$ invariant models for the superconformal gauge multiplets $\U_{\a(m) \ad(n)}$. Remarkably, we found that the functional form of the former is similar to that of \eqref{nonsusySDeqGeneral}, namely\footnote{The reader is referred to section \ref{Section3} for the appropriate definitions.}
\bea
\label{SDeqGeneral}
\ri^{m+n+1} \int \rd^{4}x \, \rd^{2\cN} \q \, {\mathcal E} \, \Big \{ \hat{\mathbb{W}}^{[\D] \a(m+n+\cN)}  \check{\mathbb{W}}^{[\D]}_{\a(m+n+\cN)}
+ \hat{\mathbb{M}}^{[\D] \a(m+n+\cN)} \check{\mathbb{M}}^{[\D]}_{\a(m+n+\cN)} \Big \} + \text{c.c.}  = 0 ~.~~~~~~~
\eea
As aforementioned, there exist more general superconformal gauge multiplets than those described by the superfields $\U_{\a(m) \ad(n)}$. Specifically, in the $\cN=2$ case, vector and superconformal gravitino multiplets do not belong to this family. In spite of this, the functional form of their self-duality equations, see  \cite{KT1,KT2,K12} for the vector multiplet and \eqref{2.13} for the gravitino multiplet,  agrees with that of \eqref{SDeqGeneral}.\footnote{There have been a number of publications devoted to solutions of the self-duality equation for the $\cN=2$ vector multiplet
\cite{KT1,KT2,K21,Carrasco:2011jv, BCFKR, CK, Ivanov:2013maa}.}
This leads us to believe that the latter is somewhat universal.

In section \ref{section2.3}, inspired by the approaches of \cite{K13,ILZ}, we described how duality-invariant models for the $\cN=2$ superconformal gravitino multiplet could be reformulated via an auxiliary superfield approach. This proved to be a powerful technique to generate such models. Below, we will sketch such a formulation for duality-invariant models describing the gauge superfield $\U_{\a(m)\ad(n)}$ described above.\footnote{One may analogously consider such a reformulation in the non-supersymmetric case, though this construction will be omitted here.}

The starting point for such an analysis is the following action functional
\begin{align}
\label{5.3}
\cS^{(m,n)}[\hat{\mathbb{W}},\check{\mathbb{W}}, \hat{\eta},\check{\eta}] =& 2 \ri^{m+n} \int \rd^4x
\rd^{2\cN} \q \, {\mathcal{E}} \, \Big \{ \hat{\eta}^{[\D] \a(m+n+\cN)}  \check{\mathbb{W}}^{[\D]}_{\a(m+n+\cN)} + \check{\eta}^{[\D] \a(m+n+\cN)} \hat{\mathbb{W}}^{[\D]}_{\a(m+n+\cN)} \non \\
&-\hat{\eta}^{[\D] \a(m+n+\cN)} \check{\eta}^{[\D]}_{\a(m+n+\cN)} - \frac{1}{2} \hat{\mathbb{W}}^{[\D] \a(m+n+\cN)} \check{\mathbb{W}}^{[\D]}_{\a(m+n+\cN)} \Big \} + \text{c.c.} \non \\
&+ \mathfrak{S}_{\text{Int}} [\hat{\eta} , \check{\eta}] ~.
\end{align}
Here we have introduced the chiral auxiliary superfields $\hat{\eta}$ and $\check{\eta}$, which are characterised by the same superconformal properties as $\hat{\mathbb{W}}$ and $\check{\mathbb{W}}$, respectively. By construction, the self-interaction $\mathfrak{S}_{\text{Int}} [\hat{\eta} , \check{\eta}]$  contains cubic and higher powers of $\hat{\eta}$, $\check{\eta}$ and their conjugates. Varying \eqref{5.3} with respect to the auxiliary superfields yields algebraic equations of motion which allow one to express $\hat{\eta}$ and $\check{\eta}$ as functions of $\hat{\mathbb{W}}$, $\check{\mathbb{W}}$ and their conjugates. This means that \eqref{5.3} is equivalent to the self-dual theory described by
\begin{align}
\cS^{(m,n)}[\hat{\mathbb{W}},\check{\mathbb{W}}] &= \ri^{m+n} \int \rd^4x \rd^{2\cN} \q \, {\mathcal{E}}\, \hat{\mathbb{W}}^{[\D] \a(m+n+\cN)} \check{\mathbb{W}}^{[\D]}_{ \a(m+n+\cN)} + \text{c.c.} + \cS_{\text{Int}} [\hat{\mathbb{W}},\check{\mathbb{W}}] ~,
\end{align}
where $\cS_{\text{Int}} [\hat{\mathbb{W}},\check{\mathbb{W}}]$ contains cubic and higher powers of $\hat{\mathbb{W}}$, $\check{\mathbb{W}}$ and their conjugates. 

This formulation is especially powerful as the self-duality equation \eqref{SDeqGeneral} takes a remarkably simple form. Specifically,  $\sU(1)$ duality invariance is equivalent to the requirement that $\mathfrak{S}_{\text{Int}}[\hat{\eta} , \check{\eta}]$ is invariant under rigid $\sU(1)$ phase transformations
\be
\mathfrak{S}_{\text{Int}} [\re^{\ri \varphi}\hat{\eta} , \re^{\ri \varphi}\check{\eta}] = \mathfrak{S}_{\text{Int}}
[\hat{\eta} , \check{\eta}] ~, \quad \varphi \in \mathbb{R} ~.
\ee

Within the Gaillard-Zumino approach to self-dual nonlinear electrodynamics
\cite{GZ1,GR1,GR2,GZ2,GZ3}, duality rotations are symmetries of the equations of motion. There exist different approaches in which duality transformations are symmetries of the action \cite{Pasti:1995tn, Pasti:2012wv, Ivanov:2014nya, Mkrtchyan:2019opf, Avetisyan:2021heg}. 
So far these approaches have not been generalised to include the (super)conformal higher-spin theories studied in our paper. The Pasti-Sorokin-Tonin (PST) formalism in four and higher dimensions (see, e.g., \cite{Pasti:1995tn, Pasti:2012wv} and references therein) has been truly successful in formulating (locally) supersymmetric theories with on-shell supersymmetry. 
It suffices to mention their construction of  
the covariant action for the super-five-brane of $M$ theory \cite{Bandos:1997ui}.
There exist several off-shell superfield generalisations of the PST formalism, see e.g. 
\cite{Pasti:1995us, Kozyrev}. It remains an interesting open problem to develop a superfield generalisations of the PST formalism for the $\sU(1)$ duality-invariant models for nonlinear $\cN=1$ supersymmetric electrodynamics   
\cite{KT1,KT2}.
\\


\noindent
{\bf Acknowledgements:} We are grateful to the referee for useful suggestions and references. SMK acknowledges kind hospitality and generous support extended to him at KIAS, Seoul, during the early stage of this project.
This work was supported in part by the Australian Research Council, project No. DP200101944 and DP230101629.

\appendix

\section{$\cN=2$ superconformal gravitino multiplet
	in $\cN=1$ superspace}\label{appendixB}

Recently, a gauge-invariant model for the free $\cN=2$ superconformal gravitino multiplet, in conformally flat backgrounds, was derived in the work \cite{HKR}. It was shown to be described in terms of a complex prepotential $\U_i$ (and its conjugate $\bar{\U}^i$), whose properties are summarised in equations \eqref{2.3} and \eqref{2.4}.
However, in the original work, the $\cN=1$ superfield content of $\U_i$ was not discussed. This appendix serves to bridge this gap.

Such an analysis requires us to work in conformally flat backgrounds, hence we set the super-Weyl tensor to zero, $W_{\a \b} = 0$. We then let $\bm{\nabla}_\a$, $\bar{\bm{\nabla}}^\ad$ and $\bm{\nabla}_\aa = \frac{\ri}{2} \{ {\bm \nabla}_\a , \bar{\bm \nabla}_\ad \}$ denote the covariant derivatives of $\cN=1$ conformal superspace \cite{ButterN=1}, see \cite{KRTM1} for a recent review. They may be defined in terms of the $\cN=2$ covariant derivatives as follows: $ {\bm \nabla}_\a {\mathfrak U} =  \nabla_\a^{\underline{1}} U|$ and $\bar{\bm \nabla}^\ad {\mathfrak U}= \bar{\nabla}^{\ad}_{\underline{1}} U|$.
Here $U$ is an $\cN=2$ superfield, and  ${\mathfrak U} \equiv U| :=U|_{\theta^\a_{\underline 2} = \bar{\theta}_\ad^{\underline 2} = 0}$ is its $\cN=1$ projection. Additionally, $\mathfrak U$ is a primary $\cN=1$ superfield if $\bold{S}^\a \mathfrak U \equiv S^{\a}_{\underline{1}} U | = 0$ and $\bar{\bold{S}}_\ad \mathfrak U \equiv \bar{S}_{\ad}^{\underline{1}} U | = 0$.

We note that the free action \eqref{2.14} is described in terms the chiral field strengths $\hat{\mathbb{W}}_\a$ and $\check{\mathbb{W}}_\a$ (and their conjugates), see eq. \eqref{2.1}. They contain the following $\cN=1$ chiral superfields in their multiplet:
\begin{subequations}
	\begin{align}
		\mathfrak{W}_{\a}^{(+)} &:= \frac{1}{\sqrt{2}} \Big( \hat{\mathbb{W}}_\a + \hf (\nabla^{\underline{2}})^2 \check{\mathbb{W}}_\a \Big) \Big| ~,  \qquad \hat{\mathfrak{W}}_{\a(2)} := \nabla^{\underline{2}}_{(\a_1} \hat{\mathbb{W}}_{\a_2)} | ~, \\
		\mathfrak{W}_{\a}^{(-)} &:= \frac{1}{\sqrt{2}} \Big( \ri \hat{\mathbb{W}}_\a - \frac{\ri}{2} (\nabla^{\underline{2}})^2 \check{\mathbb{W}}_\a \Big) \Big| ~,  \qquad \check{\mathfrak{W}}_{\a(2)} := 2\ri \nabla^{\underline{2}}_{(\a_1} \check{\mathbb{W}}_{\a_2)} | ~, \\
		\F &:= \nabla^{\a \underline 2} \check{\mathbb{W}}_{\a}|~, \qquad \qquad \qquad \qquad \quad ~
		\O_\a := \check{\mathbb{W}}_\a|~,
	\end{align}
\end{subequations}
which are primary in the sense that they are annihilated by $\bold{S}^\a$ and $\bar{\bold{S}}_\ad$. Further, owing to eq. \eqref{2.7}, the superfields $\mathfrak{W}_\a^{(\pm)}$, $\hat{\mathfrak{W}}_{\a(2)}$ and $\check{\mathfrak{W}}_{\a(2)}$ satisfy the Bianchi identities
\begin{subequations}
	\begin{align}
	{\bm \nabla}^\a {\mathfrak{W}}_\a^{(\pm)} &= \bar{{\bm \nabla}}_\ad \bar{\mathfrak{W}}^{\ad(\pm)} ~, \\
	{\bm \nabla}_\b \check{\mathfrak W}^{\a \b} &= {\bm \nabla}^{\a}{}_\bd \bar{{\bm \nabla}}_\bd \bar{\hat{\mathfrak{W}}}^{\bd(2)}~,
	\end{align}
\end{subequations}
which imply that $\mathfrak{W}_\a^{(\pm)}$ are field strengths of abelian vector multiplets, while $\hat{\mathfrak{W}}_{\a(2)}$ and $\check{\mathfrak{W}}_{\a(2)}$ are field strengths describing the $\cN=1$ superconformal gravitino multiplet. Further, $\F$ encodes the usual massless scalar multiplet, while $\O_\a$ is a chiral non-gauge superfield \cite{KPR},

These results allow us to readily reduce the free action \eqref{2.14} to $\cN=1$ superspace in accordance with the rule
\bea
\int \rd^4x\rd^4\theta\, \cE \,  \cL_c
= - \frac{1}{4} \int \rd^4x \rd^2\theta \, \bm{\cE} \, ( \nabla^{\underline{2}})^2 \cL_c| ~, \quad \bar{\nabla}^\ad_i \cL_c = 0~,
\label{A.3}
\eea
where $\bm{\cE}$ denotes the $\cN=1$ chiral measure. A routine analysis yields
\begin{align}
\cS_{\mathrm{Free}}  &= \int \rd^4 x \rd^2\q \, \bm{\cE} \, \Bigl\{ \ri
\hat{\mathfrak{W}}^{\a(2)} \check{\mathfrak{W}}_{\a(2)} + (\mathfrak{W}^{(+)})^2 + (\mathfrak{W}^{(-)})^2
+ \frac{1}{2} \F \bar{{\bm \nabla}}^2 \bar{\F} - 6 \ri \O^\a \bar{{\bm \nabla}}^2 {\bm \nabla}_\aa \bar{\O}^\ad
\Bigr\} + \text{c.c.} ~~~~~~~~~
\end{align}

\begin{footnotesize}

\end{footnotesize}

\end{document}